# Distance Measures for Sequences

Sandeep Hosangadi

*Abstract-* Given a set of sequences, the distance between pairs of them helps us to find their similarity and derive structural relationship amongst them. For genomic sequences such measures make it possible to construct the evolution tree of organisms. In this paper we compare several distance measures and examine a method that involves circular shifting one sequence against the other for finding good alignment to minimize Hamming distance. We also use run-length encoding together with LZ77 to characterize information in a binary sequence.

## 1. INTRODUCTION

The evolutionary relationship between a set of homologous characters of different organisms, taken to have descended from a common ancestral character, is represented by the phylogenetic tree. The characters can be any genic (gene sequence, protein sequence), structural (morphological) or behavioral feature of an organism and the relationship between the sequences may be expressed in terms of their mutual distance. The DNA (deoxyribonucleic acid) sequence is a string of adenine (A), guanine (G), cytosine (C) and thiamine (T) bases and there are several approaches to the characterization of the distance between such sequences. One intuitive way to characterize the distance of two equally long sequences is to see the number of places where they are different. But this may not always be the best way to characterize the distance.

In general, distance between two sequences may be measured algorithmically in terms of operations required to go from one sequence to another [1]-[6] or in terms of some geometric relationship as is done in error correction coding [7]. Distance measures are also important in linguistics and the study of scripts [8] as well as in neural networks [9],[10]. At a more fundamental level, the question of distance is related to the problem of information in physical and biological systems [11]-[17].

In a recent paper [1], the distance between two random DNA sequences was computed using the LZ data compression algorithm. In this paper, we wish to consider variations of the LZ algorithm to characterize distance and consider its relationship to LZ77 and Needleman-Wunsch algorithms [5] We compress DNA sequences using LZ77 algorithm. We will also look at the application of these ideas to binary sequences. Specifically, we will use it to associate intrinsic information with a sequence.

## 2. TYPES OF DISTANCE

Given any two sequences of equal length, the Hamming distance is the number of positions at which corresponding symbols are different. For example, consider two binary strings, 1100111001 and 0100010101. The Hamming distance is 4, because at positions 1, 5, 7 and 8, we find the symbol mismatch. The distance also works for alphabetical strings and DNA sequences.

For example if we consider two DNA sequences, AGCTAAC and AACTCCA, the Hamming distance between them will be 4, because at positions 2, 5, 6 and 7 we find symbol mismatch.

The Damerau-Levenshtein distance [6] is a distance between two sequences of finite length, where minimum transform operations are applied to transform one string into other. The transform operations include insertion, deletion or substitution of a single character or transposition of two adjacent characters. Example, if we have to calculate DL distance between CA and ABC, it takes CA→AC→ABC, where distance is 2. Here we performed two transformations, one that reverses and one insertion. However, the optimal string alignment algorithm calculates distance between any two sequences without using multiple edit operation. The distance between CA and ABC using optimal string alignment algorithm is 3 vide CA→A→AB→ABC.

The Levenshtein distance is the minimum number of changes made in spelling required to change one word into another [9].

**Example.** To calculate the Levenshtein distance between *barking* and *dark*, the following transformations have to occur:

1. barking→ barkin (deletion of g)
2. barkin→ barki (deletion of n)
3. barki→ bark (deletion of i)
4. bark→ dark (substitution of b)

Hence the Levenshtein distance between these two word strings is 4.

2.1 DICTIONARY ALGORITHMS

The information in sequences is a consequence of their patterns that can also be expressed in terms of the evolution of their probability structure. Given families of sequences, one can use Huffman coding or arithmetic coding to compress them. But when one is given a single sequence, one cannot rely on total probability. Rather, one must depend on how the probabilities of the sequence emerge as the sequence unfolds and this can be ascertained using sliding windows.

Popular universal data compression algorithms based on sliding windows were introduced by Lempel and Ziv [2],[3]. They come in two varieties: LZ77 (LZ1) and LZ78 (LZ2). These algorithms form the basis of GIF and DEFLATE algorithm used in PNG. LZ77 maintains a sliding window (length-distance pairs) during compression, whereas LZ78 uses explicit dictionary. LZ78 allows random access of input as long as Dictionary is available; however LZ77 allows sequential access of input. LZ78 algorithm does not use pre-initialized dictionary with all possible characters unlike in LZW algorithm (Lempel, Ziv, Welch [4]).

**Lempel-Ziv-Welch**

1. It encodes sequences of 8-bit data as fixed length 12 bit codes, the codes from 0 to 255 represent 1-character sequences consisting of the corresponding 8-bit character. Codes from 256 to 4095 are created in the dictionary for sequences encountered in the data as it is encoded.
2. Input bytes are grouped in a sequence till the next character makes a sequence that is not available in the dictionary.

Encoding:

1. Initialize the dictionary to contain all strings of length one.
2. Find the Longest string W in the dictionary that matches the current input.
3. Emit the dictionary index for W to output and remove W from the input.
4. Add W followed by the next symbol in the input to the dictionary.
5. Repeat Step 2.

Decoding:

1. The decoding algorithm works by reading a value from the encoded input and outputs the corresponding string from the initialized dictionary.
2. It also obtains the next value from the input and adds dictionary the concatenation of string, which was outputted with next value obtained.
3. Thus dictionary is formed at Decoding side that is exactly similar to dictionary at Encoding side.
4. Only pre-defined dictionary is exchanged between 2 parties.

## 3. ALIGNMENT ALGORITHMS

### NEEDLEMAN-WUNSCH

The Needleman-Wunsch algorithm [5] performs global alignment on two sequences, and it is used specifically in bioinformatics to align protein and nucleotide sequences. This algorithm is an example of dynamic programming, it is also referred as optimal matching Algorithm. Scores for aligned characters are specified by similarity matrix, d (Linear Gap) can be found using the similarity matrix.

The matrix $D(i,j)$ indexed by residues of each sequence is built recursively such that,

$$D(i,j) = \max\{(D(i-1,j-1)+s(x,y)), (D(i-1,j)+g), (D(i,j-1)+g)\}$$

subject to boundary conditions where $s(i,j)$ is the substitution score for residues i and j and g is the gap penalty, we construct all possible pairs of residue from two sequences.

The Needleman-Wunsch Algorithm consists of three steps:
1. Initialization of the score matrix
2. Calculation of scores and filling the trace back matrix.
3. Deducing the alignment from the trace back matrix.

**Simple Scoring Scheme:**

As example let us consider the similarity matrix below:

|   | A  | G  | C  | T  |
|---|----|----|----|----|
| A | 5  | -2 | -1 | -6 |
| G | -2 | 7  | -4 | -3 |
| C | -1 | -4 | 9  | 0  |
| T | -6 | -3 | 0  | 8  |

SEQUENCE:   AAGCTAG

CGA------G

Score with Gap penalty -5:  S(A,C)+S(A,G)+S(G,A)+(3*d)+S(G,G)

$$-1-2-2-(3*5)+7 = -13$$

There are two kind of matrices used by Needleman-Wunsch Algorithm, They are score and trace back matrices, The cells of score matrix are labeled by C(i,j) where i=1,2….N and j = 1,2…..,M.

|   |   | P   | E   | N   | D   |
|---|---|-----|-----|-----|-----|
|   | 0 | -10 | -20 | -30 | -40 |
| E | -10 |   |   |   |   |
| N | -20 |   |   |   |   |
| D | -30 |   |   |   |   |

|   |   | P    | E    | N    | D    |
|---|---|------|------|------|------|
|   |   | Left | Left | Left | Left |
| E | Up |   |   |   |   |
| N | Up |   |   |   |   |
| D | Up |   |   |   |   |

(Score Matrix)                    (Trace Back Matrix)

**Scoring:**

1. The score of any cell C(i,j) is the maximum of:

   $Qdiag = C(i-1,j-1)+S(i,j)$
   $Qup = C(i-1,j)+g$

$$Qleft = C(i,j-1)+g$$

S(i,j) is the substitution score for letters I and j and g is the gap penalty. The value of C(i,j) depends on the values of the immediately adjacent northwest diagonal, up and left cells, Once we get the score matrix we compute the traceback matrix which helps us to process deduction of the best alignment.

In the example below, we show how we compute score and trace back matrix for the words, END and PEND using BLOSUM(Block Substitution Matrix)[5].

|     | P   | E   | N   | D   |
| --- | --- | --- | --- | --- |
| 0   | -10 | -20 | -30 | -40 |
| -10 | -1  | -5  | -20 | -20 |
| -20 | -11 | -1  | 1   | -19 |
| -30 | -21 | -9  | 0   | 7   |

(Score Matrix)

|     |      | P    | E    | N    | D    |
| --- | ---- | ---- | ---- | ---- | ---- |
|     | Done | Left | Left | Left | Left |
| E   | Up   | Diag | Diag | Diag | Left |
| N   | Up   | Up   | Diag | Diag | Diag |
| D   | Up   | Diag | Diag | Diag | Diag |

(Trace back Matrix)

Traceback matrix:
1. Traceback is the process of deduction of the best alignment from the trace back matrix.
2. Traceback always begins with the last cell which is the bottom right cell.
3. It moves according to the traceback value written in the cell.
4. There are 3 possible traversals, diagonal, left or up.
5. Traceback completion happens when the top-left cell becomes "done".

Best Alignment:

1. Alignments are deduced from the values of cells which are through the traceback path, by taking into account the values of the cell that are in traceback matrix.
2. In traceback matrix, If the value is "diag", the letters from the two sequences are aligned. If it is "Left", a Gap is introduced in the left sequence, and if it is up, a gap is introduced in the top sequence, and always obtained sequences are aligned backwards.

The Needleman-Wunsch algorithm provides the best alignment over two sequences by starting its traceback from the lower right corner of the traceback matrix, and completes at the top left most cell of the matrix, it works in the similar way regardless of the length or complexity of sequences and guarantees best alignment over sequences. This Algorithm is appropriate for finding the best alignment of two sequences that are similar in length and similar across their entire lengths.

**SMITH WATERMAN**

The Smith Waterman algorithm is used to perform local alignments of sequence; it is helpful in determining identical regions between two nucleotide or protein sequences. This algorithm compares segments of all possible lengths and optimizes similarity. This algorithm uses dynamic programming to find the optimal local alignment with respect to scoring system used. Compared to Needleman-Wunsch Algorithm, this algorithm sets the negative scoring matrix cells to zero, thus positive score are only visible for backtracking, which is done by starting with highest scoring matrix cell and proceeds until a cell with zero is encountered, which gives the highest scoring local alignment.

Given two sequences A and B, Smith Waterman algorithm follows [18]:

1. The symbols in a sequence after alignment should be in same order as it was in a sequence before alignment.
2. Alignment of one symbol from one sequence with one from other is possible.
3. Alignment can be shown by a blank ('-')
4. Two blanks cannot be aligned at a time.

Scoring:

We make use of BLOSUM scoring matrix[11] for computing scores between every residue comparison.

Smith Waterman Algorithm generates N * N integer matrix, where N depicts the length of sequence taken. We compute matrix M[i][j] based on score matrix and compute optimum score.

Initially M[0][j] =0 and M[i][0]=0.

Given two sequences $A_{0--i}$ and $B_{0--j}$ we construct M[i][j] using the following rule.

1. Score        $A_i$

               $B_j$

Computation 1: $M[i-1][j-1]+ S[A_i][B_j]$

2. Gap in B     $A_i$

               $B_j$  ---

Computation 2: $M[i-1][j]-g$, where g is gap penalty.

3. Gap in A      $A_i$ --

                 $B_j$

Computation 3: M[i][j-1]-g, where g is gap penalty.

We compute all 3 computations mentioned above to find the maximum amongst them, the computation with maximum value follows its condition, for example if computation 3 holds maximum value amongst all other computation, we will be having Gap in A.

Needleman Wunsch Algorithm and Smith Waterman Algorithm make use of Gapped alignments, where we find the optimal distance between sequences by aligning them with gaps, In section 4, we discuss how we can find the optimal distance between sequences by ungapped alignments, Here we use circular shifts to find the best alignment between any two sequences.

## 4. USE OF CIRCULAR SHIFT FOR ALIGNMENT

Given two DNA sequences, we find the best alignment between them by performing circular shift by 1 bit of one sequence over the other. We find the minimum Hamming distance between any given 2 DNA sequences and conclude the distance measure between two DNA sequences.

Let AAGCTTAA and AGCTTACT be two binary sequences we try to align these sequences in such a way that we obtain minimum Hamming distance.

Sequence1: AAGCTTAA

Sequence2: AGCTTACT

We obtain a Hamming distance of 6 in the above case, If we perform circular shift of sequence 2 and again we calculate the Hamming distance,

Sequence1: AAGCTTAA

Sequence2: GCTTACTA

We obtain a Hamming distance of 7 in the above case, Likewise we hold sequence1 and perform circular shift of sequence 2 till it reaches the original sequence, we find the minimal Hamming distance obtained and assign that Hamming distance as distance measure of the given two sequences.

The Hamming distances that we get for the above two sequences after completing circular shifts of sequence 2 are:

We got 2 for last circular shift of sequence 2. i.e

Sequence 1: AAGCTTAA

Sequence2: TAGCTTAC

We got Hamming distance of 2 in this case.

We get values of 6,7,6,6,7,7,6 and 2. Since 2 is the minimal Hamming distance obtained from all circular shifts, we conclude that distance between the given two sequences is 2.

We can perform the circular shift of a sequence 2 of n bits by 1 bit, 2 bit,….n-1 bits at a time, and we can check for the minimal distance between the two sequences, that would be a better distance measure between any two DNA sequences.

## 5. COMPRESSION OF BINARY SEQUENCES

We now consider the compression of binary sequences using LZW Algorithm. We analyze the length of binary sequence before and after compression. One would normally expect the compressed sequence to be shorter than the original sequence. But as we see, this is not always true and, therefore, LZW does not work very well for binary sequences.

Table 1. Compression of random binary sequences using LZW

| Prime Numbers | Length of Binary String Before Compression | Length of Binary String After Compression |
| --- | --- | --- |
| 3 | 0 | 1 |
| 5 | 2 | 3 |
| 7 | 4 | 5 |
| 11 | 6 | 9 |
| 13 | 10 | 10 |
| 17 | 12 | 14 |
| 19 | 16 | 21 |
| 23 | 18 | 24 |
| 29 | 22 | 29 |
| 31 | 28 | 28 |
| 37 | 30 | 44 |
| 41 | 36 | 39 |
| 43 | 40 | 47 |
| 47 | 42 | 54 |
| 53 | 46 | 65 |
| 59 | 52 | 65 |
| 61 | 58 | 74 |
| 67 | 60 | 77 |
| 71 | 66 | 76 |
| 73 | 70 | 67 |
| 79 | 72 | 89 |
| 83 | 78 | 96 |
| 89 | 82 | 97 |
| 97 | 88 | 114 |

We considered several compression of binary D-sequences [19],[20],[21], which are pseudorandom sequences that represent the reciprocal of prime numbers. The D-sequence in base 10 for the prime 7 is 142857, which is the decimal expansion of 1/7. The binary D-sequence for the prime 19 is 000011010111100101. The period of the maximum-length D-sequence is one less than the prime. The table below gives the length of D binary sequences before and after LZW compression is applied.

We also tried to compare the Hamming distance between two binary sequences before and after applying LZW compression; of the two binary sequences that we considered, one is generated randomly using pseudo-random generator, and the other is a binary sequence obtained by D sequence of a prime number.

Table 2.

| Prime Number | Before LZW Compression | After LZW Compression |
|---|---|---|
| 2 | 0 | 0 |
| 3 | 0 | 0 |
| 5 | 2 | 1 |
| 7 | 4 | 3 |
| 11 | 5 | 4 |
| 13 | 4 | 11 |
| 17 | 8 | 10 |
| 19 | 7 | 9 |
| 23 | 12 | 9 |
| 29 | 13 | 17 |
| 31 | 11 | 14 |
| 37 | 22 | 21 |
| 41 | 25 | 23 |
| 43 | 23 | 27 |
| 47 | 22 | 30 |
| 53 | 27 | 36 |
| 59 | 30 | 33 |
| 61 | 35 | 35 |
| 67 | 32 | 47 |
| 71 | 35 | 34 |
| 73 | 35 | 43 |
| 79 | 40 | 47 |
| 83 | 44 | 54 |
| 89 | 40 | 44 |
| 97 | 40 | 57 |

We analyze the distance measures considered by Otu and Sayood in [1] to compute the distance between any two phylogenetic sequences. Let us consider three DNA sequences S, R and Q, and we will compute distance SQ, RQ and QS using the algorithm suggested by them.

**S** – AACGTACCATTG    **R-** CTAGGGACTTAT

**Q-** ACGGTCACCAA

We compute history H(S), H(R), H(Q), H(SQ), H(RQ) and H(QS). Also we compute C(S), C(R), C(Q), C(SQ), C(RQ) and C(QS).

**H(S)** – A. AC.G.T.ACC.AT.TG        **C(S)** – 7

**H(R)-** C.T.A.G.GGA.CTT.AT        **C(R)** – 7

**H(Q)-** A. C.G.GT.CA.CC.AA        **C(Q)-** 7

**H(SQ)-** A.AC.GT.ACC.AT.TG.ACGG.TC.ACCAA    **C(SQ)-** 10

**H(RQ)-** C.T.A.G.GGA.CTT.AT.ACG.GT.CA.CC.AA  **C(RQ)-** 12

**H(QS)** – A.C.G.GT.CA.CC.AA.AAC.GTA.CCAT.TG    **C(QS)-** 11

Now we compute the distance measures as suggested in [1].

1. d(S,Q) = max{C(SQ)-C(S), C(QS)-C(Q)}

    max{10-7, 11-7} = 4

2. d*(S,Q) = {C(SQ)-C(S)+C(QS)-C(Q)}

    {10-7+11-7} = 7

We are looking for a new way of calculating components for any given binary sequence, for example let us consider a binary sequence 000110101111100101, We can calculate the number of components using Lempel Ziv's strategy. In LZ Strategy, A binary sequence is scanned from left to right. For every new String it encounters, it adds up the String into the dictionary. If it encounters a string that is already present in the dictionary, it performs Run-Length coding discussed in the below example.

i.  0| 000 | 1| 10| 101| 11| 100| 101

ii. 0| 30 |1| 00 | 01 | 10| 11 | 100

Number of components can be calculated is 8 because we can break the binary sequence into 8 components as shown in (i). The dictionary stores the exact values at appropriate indices as shown below:

| Index | String |
|-------|--------|
| 0     | 0      |
| 1     | 1      |
| 00    | 10     |
| 01    | 101    |
| 10    | 11     |
| 11    | 100    |
| 100   | 101    |

In the second component as shown in (i), we make use of Run-Length coding. Since 0 is repeated for 3 times, we represent 000 as 30. Run length coding is effective when a component have symbols that have more than 3 repetitions. Given a binary sequence, we try to calculate number of components that can be built using LZ77 algorithm. We explain the concept by an example. Let us consider a binary sequence 0010101011011101. We show how we divide this sequence into components using LZ77 technique with and without implementation of Run-length encoding.

With Run-length encoding:

**0| 01 |010101 |1| 011 |10 |1**

No of Information Units → 7

Without Run-length encoding:

**0| 01 |010 |1 |011 |0111| 01**

No of Information Units → 7

**Binary sequence 2:** 000011010111100101

With Run-length encoding:

**0| 000 | 1| 10| 101| 11| 100| 101**

No of Information Units → 8

Without Run-Length encoding:

**0 |00 |01 |1 |010| 11 |110 |0101**

No of Information Units → 9

**ALGORITHM (WITH RUNLENGTH ENCODING):**

1. Read the given binary sequence Left to Right with single character at a time.
2. If the current character is not present in the dictionary, add the character to dictionary, Add to component list, Move to next character.
3. If current character is present in the dictionary, Buffer up the next characters till the longest substring match is found with dictionary words.
4. Perform Run-length coding, Also concatenate the string till completion of Run-length encoding, Add the final string to component list.
5. Calculate Number of components and display component array.

In the pseudo-code written below, there are four methods LONGESTSUBSTRINGSEARCH(int), SEARCHDICTIONARY(String). COMPONENT(String), DICTIONARY(String).

Amongst these, the first one figures out the longest substring match of buffered sequences with obtained dictionary and returns the longest string. The second one searches the string in dictionary, if string was found in dictionary, it returns 1 else returns 0. The third and the fourth ones store the string into COMPONENT and DICTIONARY arrays.

PSEUDOCODE:

BEGIN

    WHILE (i<BINARYSTRING.LENGTH)

        STRING S1 ←BINARYSTRING.CHARAT(i)

    IF(SEARCHDICTIONARY(S1))

        LONGEST ←LONGEST SUBSTRINGSEARCH(i)

        LONGLEN ← LONGEST.LENGTH();

        IF(LONGEST.EQUALS(NEXTBUFFEREDSEQUENCE)

            WHILE (LONGEST.EQUALS(NEXTBUFFEREDSEQUENCE)

                PERFORM RUNLENGTH ENCODING

                I ←I+LONGLEN

        ELSE

            WHILE(CONCATENATED STRING NOT IN DICTIONARY)

                DO CONCATENATION

            ADD THE CONCATENATED STRING TO DICTIONARY

ADD THE CONCATENATED STRING TO COMPONENT

END

## ALGORITHM WITHOUT IMPLEMENTATION OF RUN LENGTH ENCODING:

1. Scan the sequence from Left to Right character by character.

2. Check the shortest string not in Dictionary by concatenating existing string encountered. add it to the component array. Move to next character.

3. Repeat Step 2 till we reach end of sequence.

PSEUDOCODE:

WHILE(BINARYSEQUENCE.LENGTH()>I)

    S1←BINARYSEQUENCE.CHARAT(I)

    IF SEARCHDICTIONARY(S1)

        WHILE SEARCHDICTIONARY(S1)

            S1 =S1.CONCAT(CHARAT(I))

            I++

        END WHILE

        COMPONENT(S1)

        DICTIONARY(S1)

    ELSE

        COMPONENT(S1);

        DICTIONARY(S1);

    END IF ELSE

END WHILE

DISPLAY COMPONENTS

## 5. COMPARISION

Given any sequence, If a sequence is more random, number of components calculated within the sequence using Run-length encoding is almost equal to one without using Run-length encoding,

Otherwise, if sequence is less random Run-length encoding works better for component calculation within the sequence.

We try to plot a graph of sequence length against component size when a given sequence is more and less random.

We say a sequence is less random when a sequence contains more successive strings that are superset of previous encountered strings. If not, we say they are more random.

When a sequence is less random, use of run-length coding helps improve the performance substantially as shown below.

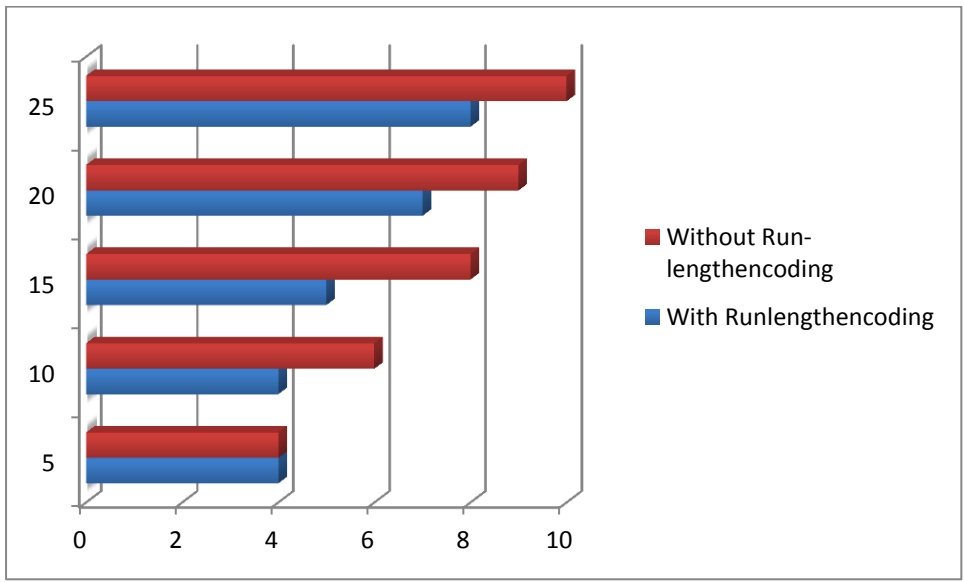

When a sequence is more random the results are not much different as shown by the graph below.

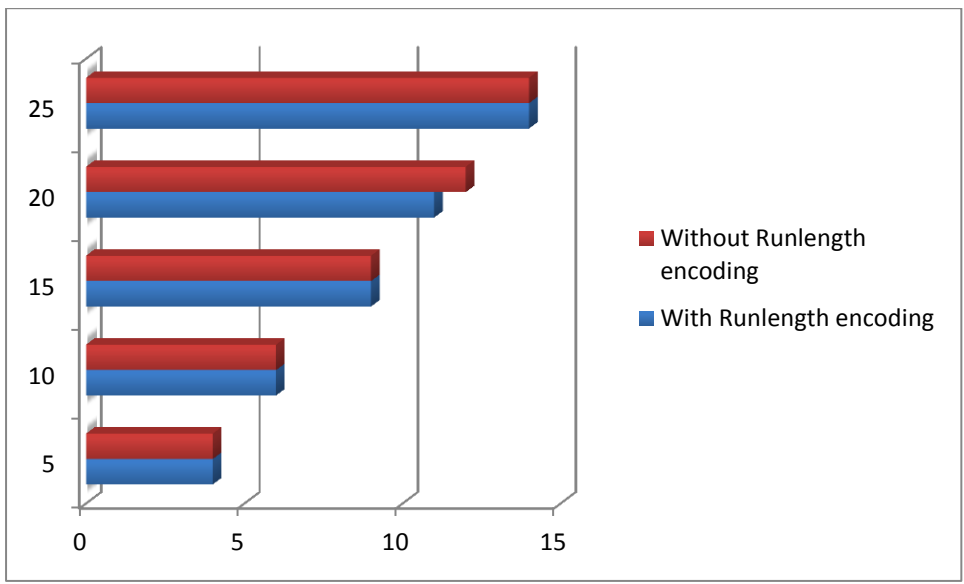

# 6. CONCLUSION

The paper has examined ideas on distance measures that can apply both to genomic sequences and to more abstract binary sequences. We proposed that in some situations consideration of circular shift can be helpful in finding the minimum distance between any two sequences. We also discussed the component size of any given sequence with and without implementation of run-length encoding. Further issues related to statistical significance [21]-[26] will be considered in a companion paper.

## REFERENCES

[1] H.H. Otu and K. Sayood, A new sequence distance measure for phylogenetic tree construction. Bioinformatics 19: 2122-2130, 2003.
[2] J. Ziv and A. Lempel, A Universal Algorithm for Sequential Data Compression. IEEE Trans. Information Theory IT-23: 337-343, 1977.
[3] J. Ziv and A. Lempel, "Compression of Individual Sequences via Variable-Rate Coding. IEEE Trans. Information Theory, Vol. IT-24: 5306, 1978.
[4] T.A. Welch, A technique for high-performance data compression. IEEE Computer, 1984.
[5] S.B. Needleman and C.D. Wunsch, A general method applicable to the search for similarities in the amino acid sequence of two proteins. Journal of Molecular Biology 48: 443–53, 1970.
[6] D. Gusfield, Algorithms on strings, trees, and sequences: computer science and computational biology. Cambridge University Press, 1997.
[7] T.K. Moon, Error correction coding. Wiley-Interscience, 2005.
[8] S. Kak, A frequency analysis of the Indus script. Cryptologia 12: 129-143, 1988.
[9] S. Kak, A class of instantaneously trained neural networks. Information Sciences 148: 97-102, 2002.
[10] S. Kak, Self-indexing of neural memories, Physics Letters A 143: 293-296, 1990.
[11] R. Landauer, The physical nature of information. Physics Letters A 217: 188-193, 1996.
[12] S. Kak, Information, physics and computation. Foundations of Physics 26: 127-137, 1996.
[13] S. Kak, Quantum information and entropy. International Journal of Theoretical Physics 46: 860-87, 2007.
[14] A. Kolmogorov, Three approaches to the quantitative definition of information. Problems of Information Transmission. 1: 1-17, 1965.
[15] S. Kak, Classification of random binary sequences using Walsh-Fourier analysis. IEEE Trans. Electronic Compatibility EMC-13, 1971.
[16] M. .A. Nielsen and I.L. Chuang, Quantum Computation and Quantum Information. Cambridge University Press, 2000.

[17] S. Kak, The initialization problem of quantum computing. Foundations of Physics 29: 269-279, 1999.

[18] S. F. Altschul and B. W. Erickson, Optimal sequence alignment using affine gap costs. Bulletin of Mathematical Biology 48: 603–616, 1986.

[19] S. Kak and A. Chatterjee, On decimal sequences. IEEE Transactions on Information Theory IT-27: 647-652, 1981.

[20] S. Kak, Encryption and error-correction coding using D sequences. IEEE Transactions on Computers C-34: 803-809, 1985.

[21] S. Kak, New results on d-sequences. Electronics Letters 23: 617, 1987.

[22] A. Agrawal and X Huang, Pairwise statistical significance of local sequence alignment using sequence specific and position specific substitution matrices. IEEE Trans on Comp. Biology and Bioinformatics 8: 194-205, 2011.

[23] M. Kschischo, M. Lässig, and Y.-K. Yu, Toward an Accurate Statistics of Gapped Alignments. Bull. of Math. Biology 67: 169-191, 2004.

[24] S. Sheetlin, Y. Park, and J.L. Spouge, The Gumbel Pre-Factor k for Gapped Local Alignment Can Be Estimated from Simulations of Global Alignment. Nucleic Acids Research 33: 4987-4994, 2005.

[25] A. Y. Mitrophanov and M. Borodovsky, Statistical Significance in Biological Sequence Analysis. Briefings in Bioinformatics 7: 2-24, 2006.

[26] Y.-K. Yu and S.F. Altschul, The Construction of Amino Acid Substitution Matrices for the Comparison of Proteins with Non-Standard Compositions. Bioinformatics 21: 902-911, 2005.